\documentclass[5p]{elsarticle} 

\usepackage{lineno}
\modulolinenumbers[5]
\usepackage{mathtools}
\journal{ Prog. in Photov. Special Issue 36$^{th}$ EU PVSEC}

\usepackage{eurosym}
\usepackage{threeparttable} 

\usepackage{url}
\usepackage[colorlinks=true, citecolor=blue, linkcolor=blue, filecolor=blue,urlcolor=blue]{hyperref}

\usepackage{lineno,hyperref}
\modulolinenumbers[1]
\usepackage{amsmath}
\usepackage{siunitx}
\biboptions{numbers,sort&compress}
\usepackage[europeanresistors,americaninductors]{circuitikz}
\usepackage{adjustbox}
\usepackage{xspace}
\usepackage{caption}
\usepackage{booktabs}
\usepackage{tabularx}
\usepackage{threeparttable}
\usepackage{multicol}
\usepackage{float}
\usepackage{graphicx,dblfloatfix}
\usepackage{csvsimple}

\def\el{${}_{\textrm{el}}$}
\def\th{${}_{\textrm{th}}$}

\begin{document}

\begin{frontmatter}

\title{The role of photovoltaics in a sustainable European energy system under variable CO$_2$ emissions targets, transmission capacities, and costs assumptions}

\author[mymainaddress,iClimate]{Marta Victoria\corref{mycorrespondingauthor}}
\ead{mvp@eng.au.dk}
\author[mymainaddress]{Kun Zhu}
\author[kitaddress]{Tom Brown}
\author[mymainaddress,iClimate]{Gorm B. Andresen}
\author[mymainaddress,iClimate]{Martin Greiner}
\cortext[mycorrespondingauthor]{Corresponding author}
\address[mymainaddress]{Department of Engineering, Aarhus University, Inge Lehmanns Gade 10, 8000 Aarhus, Denmark}
\address[iClimate]{iCLIMATE Interdisciplinary Centre for Climate Change, Aarhus University}
\address[kitaddress]{Institute for Automation and Applied Informatics (IAI), Karlsruhe Institute of Technology (KIT), Forschungszentrum 449, 76344, Eggenstein-Leopoldshafen, Germany}

\begin{abstract}
PyPSA-Eur-Sec-30 is an open-source, hourly-resolved, networked model of the European energy system which includes one node per country as well as electricity, heating, and transport sectors. The capacity and dispatch of generation and storage technologies in every country can be cost-optimised under different CO$_2$ emissions constraints. This paper presents an overview of the most relevant results previously obtained with the model, highlighting the influence of solar photovoltaic (PV) generation on them. For 95\% CO$_2$ emissions reduction, relative to 1990 level, PV generation supplies in average 33\% of the electricity demand. Southern European countries install large PV capacities together with electric batteries, while northern countries install onshore and offshore wind capacities and use hydrogen storage and reinforced interconnections to deal with wind fluctuations. The strong daily generation profile of solar PV heavily impacts the dispatch time series of backup generation and storage technologies. The optimal PV and wind capacities are investigated for variable CO$_2$ emissions, transmission interconnection among neighbouring countries, and cost assumptions. 

\end{abstract}

\end{frontmatter}


\section{Introduction}
\label{sec_introduction}

The European Commission has recently called for a climate-neutral Europe by 2050 \cite{in-depth_2018}. This aligns with the EU commitment, through the Paris Agreement \cite{Paris_agreement}, to keep the global temperature rise this century, well below 2 $^{\circ}$C above pre-industrial levels and to pursue efforts to limit temperature increase even further to  1.5 $^{\circ}$C. The energy supply accounted for one quarter of the EU CO$_2$ emissions in 2016 \cite{GHG_by_sector}. Moreover, significant shares of the greenhouse gas emissions in the Transport, Industry, and Residential and Commercial sectors, which represented 18\%, 17\%, and 11\%, respectively, are related to energy provision \cite{GHG_by_sector}. Solar photovoltaics (PV), thanks to its dramatic cost decrease during the last decade \cite{Louwen_2016, Haegel_2017}, has transformed from being a niche technology used to power expensive satellites or remote isolated power systems, to be considered one of the key technologies that could enable a timely transition to a decarbonised Europe. Almost every actor in the energy sector has underestimated the potential growth of solar PV. Among others, the International Energy Agency (IEA) \cite{Fell_2015, Haegel_2017}, Greenpeace \cite{Creutzig_2017}, and even PV scientists \cite{Haegel_2019}. The reality today is that global PV installed capacity exceeded 500 GW at the end of 2018, and a significant expansion is expected in the coming years \cite{Haegel_2019}. \\ 

In a future energy system that relays mostly on solar and wind energy, the fluctuating nature of these sources requires balancing their generation. Three main strategies are described in the literature to that end: the deployment of storage technologies, the geographical integration of variable generation trough interconnections among neighbouring countries, and the local integration through sector coupling that brings additional flexibility to the system. \\

Previous efforts to investigate highly renewable penetration in Europe include modelling the networked power system using weather-driven models \cite{Heide_2010, Heide_2011, Rasmussen_2012, Rodriguez_2014, Eriksen_2017, Raunbak_2017}, rule-based models \cite{Bussar_2016}, or cost optimisation \cite{Schlachtberger_2017, Gils_2017a, Cebulla_2017, Collins_2018}; simulating the sector-coupled European energy system under the copperplate assumption \cite{Connolly_2016}; and modelling the transition path to a low-carbon European energy system \cite{ Breyer_2017, Breyer_2017EUPVSEC, Child_2019}. A comprehensive list of the key findings in all the relevant models for Europe is provided in \cite{Child_2019}. For detailed discussions on issues regarding the technical feasibility of 100\% renewable-electricity systems, \cite{Brown_response} is recommended.\\

In this work, we use the PyPSA-Eur-Sec-30 model that combines the three mentioned renewable-balancing strategies to investigate the contribution of solar PV to a future highly renewable European energy system. The model has been used to investigate the synergies of sector coupling and transmission reinforcement \cite{Brown_2018}, the impact of CO$_2$ prices on the coupled heating-electricity European system \cite{Zhu_2019}, the system properties as the CO$_2$ emissions are restricted \cite{Brown_2019}, and the role of different storage technologies \cite{Victoria_storage}. Furthermore, the European power system has been analysed in \cite{Schlachtberger_2017, Schlachtberger_2018}.  Here we focus on investigating the optimal PV penetration and how does solar generation impact the required capacities and dispatch patterns of other generation and storage technologies. \\

The paper is organised as follows. Section \ref{sec_methods} includes a brief description of the model and data used. Section \ref{sec_results} gathers the main results previously obtained with PyPSA-Eur-Sec-30 highlighting the influence of solar PV penetration on them. Moreover, a sensitivity analysis to evaluate the impact of CO$_2$ emissions cap, transmission expansion, and cost assumptions is included. Finally, the main conclusions are summarised in Section \ref {sec_conclusions}.  
\section{Methods} \label{sec_methods}

\paragraph{Model}\

\begin{figure}[!h]
\centering
	\includegraphics[width=\columnwidth]{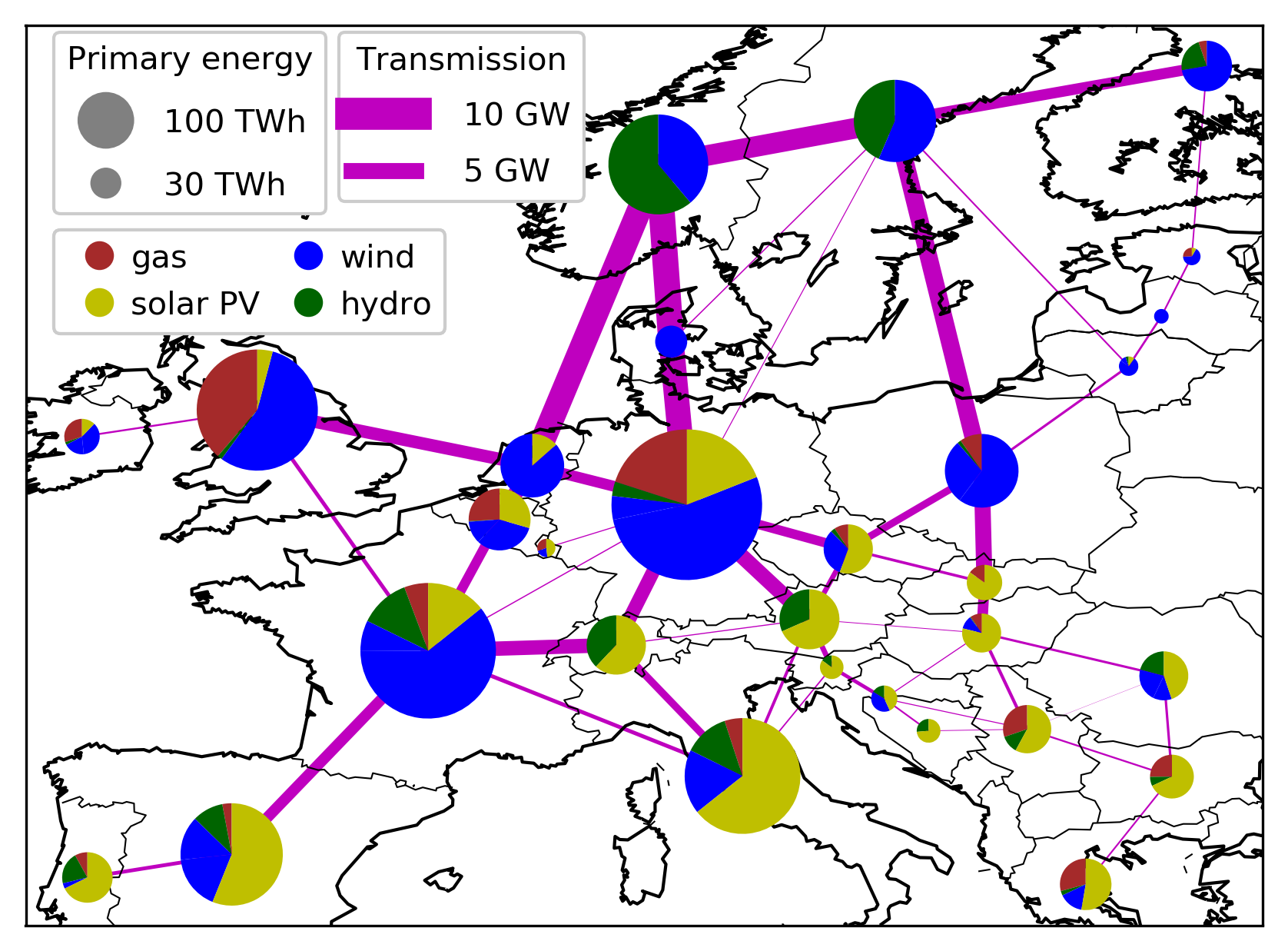}
\caption{Primary energy in every country for the optimal power system configuration under 5\% CO$_2$ emissions constraint, relative to 1990 level, and maximum transmission expansion limited to twice today's volume.} \label{fig_spatial_plot} 
\end{figure}

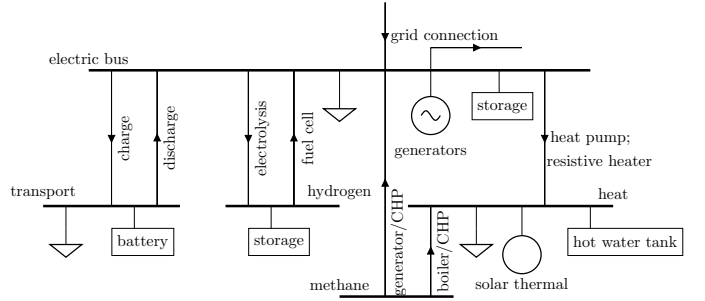
\begin{figure}[!t]
  \begin{adjustbox}{scale=0.60,trim=5 6.8cm 0 0}
  \begin{circuitikz}
  \draw (1.5,14.5) to [short,i^=grid connection] (1.5,13);
  \draw [ultra thick] (-5,13) node[anchor=south]{electric bus} -- (6,13);
  \draw(2.5,13) |- +(0,0.5) to [short,i^=$$] +(2,0.5);
  \draw (0,-0.5) ;
  \draw (0.5,13) -- +(0,-0.5) node[sground]{};
  \draw (2.5,12) node[vsourcesinshape, rotate=270](V2){}
  (V2.left) -- +(0,0.6);
  \draw (2.5,11.2) node{generators};
    \node[draw,minimum width=1cm,minimum height=0.6cm,anchor=south west] at (3.4,11.9){storage};
    \draw (4,13) to (4,12.5);

  \draw [ultra thick] (-6,10) node[anchor=south]{transport} -- (-3,10);
  \draw (-5.5,10) -- +(0,-0.5) node[sground]{};
  \draw (-3.5,10) to [short,i_=${}$] (-3.5,13);
  \draw (-3.2,11.5)  node[rotate=90]{discharge};
  \draw (-4.5,13) to [short,i^=${}$] (-4.5,10);
  \draw (-4.2,11.5)  node[rotate=90]{charge};
  \node[draw,minimum width=1cm,minimum height=0.6cm,anchor=south west] at (-4.5,8.9){battery};
  \draw (-4,10) to (-4,9.5);

    \draw [ultra thick] (2,10) -- (6.5,10)  node[anchor=south]{heat};
  \draw (3.5,10) -- +(0,-0.5) node[sground]{};
  \draw (4.5,9.35) to [esource] (4.5,8.5);
  \draw (4.5,10) -- (4.5,9.35);
  \draw (4.5,8.3) node{solar thermal};
  \draw (5,13) to [short,i^=heat pump;] (5,10);
  \draw (6.2,11) node{resistive heater};
  \node[draw,minimum width=1cm,minimum height=0.6cm,anchor=south west] at (5.5,8.9){hot water tank};
  \draw (6,10) to (6,9.5);

  \draw [ultra thick] (-2,10)  -- (0.5,10) node[anchor=south]{hydrogen};
  \draw (-1.5,13) to [short,i_=${}$] (-1.5,10);
    \draw (-1.2,11.5)  node[rotate=90]{electrolysis};
  \draw (-0.5,10) to [short,i^=${}$] (-0.5,13);
  \draw (-0.2,11.5)  node[rotate=90]{fuel cell};
  \draw (-1,10) to (-1,9.5);
  \node[draw,minimum width=1cm,minimum height=0.6cm,anchor=south west] at (-1.5,8.9){storage};

  \draw [ultra thick] (0.5,8) node[anchor=south]{methane} -- (3,8);
  \draw (1.5,8) to [short,i_=${}$] (1.5,13);
  \draw (2.5,8) to [short,i_=${}$] (2.5,10);
  \draw (1.8,9.2)  node[rotate=90]{generator/CHP};
  \draw (2.8,9)  node[rotate=90]{boiler/CHP};
  \end{circuitikz}
\end{adjustbox}
\caption{Energy flow at a single node representing a country. Within each node there is a bus (thick horizontal line) for every sector (electricity, transport and heating), to which different loads (triangles), energy sources (circles), storage units (rectangles) and converters (lines connecting buses) are attached.}
\label{Fig_buses}
\end{figure}

The PyPSA-Eur-Sec-30 model is an open-source, hourly-resolved, one-node-per-country network model of the European energy system. It currently includes the electricity, heating, and transport sectors \cite{Brown_2018}. The network, shown in Figure \ref{fig_spatial_plot}, comprises 30 nodes, the 28 European Union member states as of 2018 excluding Malta and Cyprus but including Norway, Switzerland, Serbia, and Bosnia-Herzegovina. Neighbouring countries are connected through High Voltage Direct Current (HVDC) transmission lines, whose capacities can be expanded if it is cost-effective. Figure \ref{Fig_buses} shows the three buses representing the electricity, heating, and transport sectors in every country.

The capacity and dispatch of every asset (generation, storage, and transmission capacities) are jointly optimised assuming perfect foresight and competition as well as long-term market equilibrium. The model is built in the framework Python for Power System Analysis (PyPSA) \cite{PyPSA} and other instances exist with higher spatial resolution \cite{Horsch_2018, Horsch_2017}. The optimal system configuration is determined by minimising the total annualised system cost calculated as:

\begin{align}
& \min_{\substack{G_{n,s},E_{n,s},\\F_\ell,g_{n,s,t}}} \left[ \sum_{n,s} c_{n,s} \cdot G_{n,s} +\sum_{n,s} \hat{c}_{n,s} \cdot E_{n,s} \right. \nonumber \\
& \hspace{2cm} \left. + \sum_{\ell} c_{\ell} \cdot F_{\ell}+ \sum_{n,s,t} o_{n,s,t} \cdot g_{n,s,t} \right]
\label{eq_objective}
\end{align}

where $c_{n,s}$ are the fixed annualised costs for generator and storage power capacity $G_{n,s}$ of technology $s$ in every bus $n$, $\hat{c}_{n,s}$ are the fixed annualised costs for storage energy capacity $E_{n,s}$, $c_\ell$ are the fixed annualised costs for bus connectors $F_{\ell}$, and  $o_{n,s,t}$ are the variable costs, for generation and storage dispatch $g_{n,s,t}$  in every hour $t$. Bus connectors $\ell$ include transmission lines but also converters between the buses implemented in every country to represent the different sectors, for instance, heat pumps that connect the electricity and heating bus. The optimisation of the system is subject to several constraints, such as ensuring the supply of inelastic demand in every sector, a limited expansion of the transmission capacities or a maximum CO$_2$ emissions $\textrm{CAP}_{CO2}$. For example, the latter is imposed by

\begin{equation}
  \sum_{n,s,t}  \varepsilon_{s} \frac{ g_{n,s,t} }{\eta_{n,s}}  \leq  \textrm{CAP}_{CO2} \hspace{.4cm} \leftrightarrow \hspace{0.3cm} \mu_{CO2} \label{eq_co2cap}
\end{equation}
where $\varepsilon_{s}$ represents the specific emissions in CO$_2$-tonne-per-MWh\th{} of the fuel $s$, $\eta_{n,s}$ the efficiency and $g_{n,s,t}$ the generators dispatch. The Lagrange/Karush-Kuhn-Tucker multiplier $\mu_{CO2}$ represents the CO$_2$ shadow price, \textit{i.e.}, the required CO$_2$ price to achieve the emissions reduction in an open market. \\

Greenfield optimisation is implemented, \textit{i.e.}, the European energy system is built from scratch. Cost assumptions correspond to projections for 2030 for all technologies. This time horizon allows including the forecasted cost decrease, mostly for wind, solar PV and storage, while avoiding large uncertainties associated with long-term cost projections. Electricity can be produced by onshore and offshore wind, solar PV, Open-Cycle Gas Turbines (OCGT), Combined Heat and Power (CHP) plants, and hydro power plants. The capacity of the latter is exogenously fixed and considered fully amortised. Electricity can be stored in static electric batteries, overground hydrogen storage, and Pump Hydro Storage (PHS). Coupling heating and transport sectors brings new generation and storage technologies to the system, which for the sake of conciseness are not described here. A detailed description of the model can be found in \cite{Brown_2018} and \cite{Victoria_storage}.

\paragraph{Code} \

The PyPSA-Eur-Sec-30 model is available through the repository \href{https://doi.org/10.5281/zenodo.1146666}{10.5281/zenodo.1146666}. Moreover, code to plot the figures shown in this paper is available at \newline
\href{https://github.com/martavp/PyPSA-plots.git}{github.com/martavp/PyPSA-plots.git}. 

\paragraph{Data}\

The efficiencies, lifetimes, and costs assumptions are summarised in Table \ref{tab_costs}. A thorough description of all the data used can be found in \cite{Brown_2018, Zhu_2019, Victoria_storage}. Time series corresponding to 2015 are used for the demands, as well as solar and wind generation, profiles. For instance, historical data is used to represent the electricity demand and the population-weighted temperature time series are used to convert annual values of heat demand into hourly resolution. 

\begin{table*}[!t]
\centering
	\begin{threeparttable}
		\caption{Costs, lifetime, and efficiency values assumed in the model. The table only shows technologies in the power sector, details for the technologies in the heating and trasport sectors can be found in \cite{Brown_2018}.} \label{tab_costs}
		\centering
		\begin{footnotesize}
		\begin{tabularx}{0.8\textwidth}{lcccccc}
			\toprule
			Technology                 				&Overnight   						&Unit 	&FOM\tnote{b} 	&Lifetime 		&Efficiency 		&Source		\\
																				&Cost\tnote{a}[\euro] 	&     	&[\%/a] 				&[years]   	  &   						& 				\\
			\midrule
			Onshore wind             							&910     						&kW\el  &3.3 						&30  					&	    					& \cite{DEA_2016}\\
			Offshore wind             						&2506        				&kW\el  &3   						&25  &				& \cite{DEA_2016} \\
			Solar PV utility-scale\tnote{c}  			&425         				&kW\el  &3 							&25  &	 			&  \cite{Vartiainen_2017} \\
			Solar PV rooftop\tnote{c}  		  		&725         				&kW\el  &2 							&25  &	 			&  \cite{Vartiainen_2017} \\
			OCGT\tnote{d}              
									&560         				&kW\el  &3.3 						&25  &0.39 		&  \cite{DEA_2016,Schroeder_2013} \\
			Hydro reservoir\tnote{e} & 2000 & kW\el & 1 & 80  & 0.9 & \cite{Schroeder_2013} \\
			Run-of-river\tnote{e} & 3000 & kW\el & 2 & 80 & 0.9 & \cite{Schroeder_2013}  \\
			Pumped hydro storage\tnote{e} (PHS)            & 2000 							&KW\el 	& 1     				&80 &0.87$\cdot$0.87=0.76 & \cite{Schroeder_2013}\\
      Batteries   													& 144.6 						&KWh\el & 0    &15 &0.9$\cdot$0.9=0.81\tnote{f} & \cite{Budischak_2013}\\
      Battery inverter  										&310                &KW\el  &3     &20 &0.9\tnote{f}  & \cite{Budischak_2013} \\
      Hydrogen storage\tnote{g}                      &8.4                &KWh\el &0     &20 &0.8$\cdot$0.58=0.46 &\cite{Budischak_2013}\\
      Hydrogen electrolysis  								&350 								&KW\el  &4     &18 &0.8 & \\
      Hydrogen fuel cell                   &339 								&KW\el  &3     &20 &0.58& \cite{Budischak_2013}\\  
			HVDC lines	  			   &400			&MWkm	&2	 &40  &		   1			&  \cite{Hagspiel_2014}						\\
			HVDC converter pair	  			   &150			&kW	&2	 &40  &		1   			& \cite{Hagspiel_2014}						\\
			\bottomrule
		\end{tabularx}
		\end{footnotesize}
		\begin{tablenotes}
			\footnotesize
			\item [a] Costs are annualised assuming a discount rate of 0.07.
			\item [b] Fixed Operation and Maintenance (FOM) costs are given as a percentage of the overnight cost per year.
			\item [c] 50\% of the installed capacities are rooftop-mounted systems and 50\% utility-scale power plants, 4\% and 7\% discount rate has been assumed, respectively. 
			\item [d] The fuel cost of Open-Cycle Gas Turbines (OCGT) is 21.6 \EUR/MWh$_{th}$. 
			\item [e] Reservoir, run-of-river and PHS are exogenous to the system. The capacities in every country are fixed and they are considered to be fully amortised
      
			\item [f] A conservative value of 90\% has been assumed for the charging and discharging efficiency of batteries, but slightly 		 higher values are already attained \cite{Weniger_2016}.
			\item [g] Cost for overground steel tanks is assumed here but underground storage of hydrogen can be significantly cheaper \cite{Steward_2009}. 
		\end{tablenotes}
	\end{threeparttable}
	\centering
\end{table*}

\begin{figure}[ht!]
\centering
\includegraphics[width=\columnwidth]{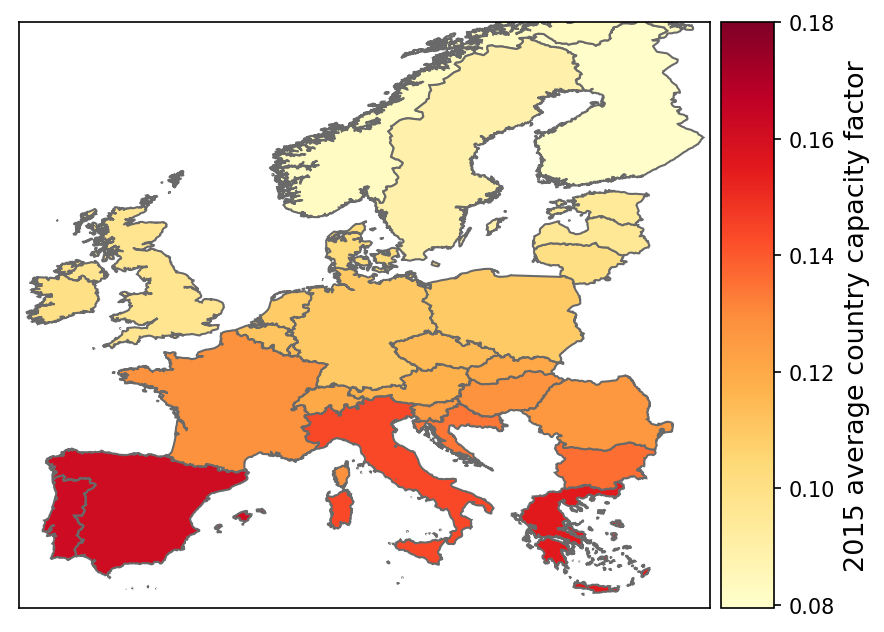}\\
\caption{Country-wise annual capacity factors estimated assuming uniform distributions of rooftop installations in every country. A detailed description of the methodology to generate the time series can be found in \cite{Victoria_2019b}. } \label{fig_map}
\end{figure}

\paragraph{Time series representing PV at national scale}\

The Global Renewable Energy Atlas (REatlas) \cite{Andresen_2015} is used to obtain time series representing the hourly capacity factor in every country. Irradiance from Climate Forecast System Reanalysis (CFSR) \cite{CFSR} is used as input. CFSR dataset has a spatial resolution of 0.3125$^{\circ}$x0.3125$^{\circ}$, which corresponds to roughly 40x40 km$^2$ in Europe. Since reanalysis irradiance is known to include significant bias, CFSR irradiance is first corrected using SARAH satellite dataset. Then, the bias-corrected global irradiance is split into direct and diffuse irradiance. The components direct, diffuse, and albedo on a tilt panel are computed using an anisotropic sky model and aggregated to obtain the irradiance at the PV panel aperture. A temperature-dependent efficiency model is used to transform irradiance to PV generation in every CFSR grid cell. The bias-correction allows using reanalysis irradiance, so that time series are consistent with those obtained for wind and hydroelectricity, while keeping uncertainty similar to that of satellite irradiance.\\

To obtain the time series representative for PV generation in every country, hourly values of irradiance and temperature in every grid cell (40x40km$^2$) are converted into PV generation and grid cells within a country are aggregated. In order to represent the behaviour of a myriad of small PV installations scattered across every country, a constant capacity layout, \textit{i.e.}, constant PV capacity per grid cell is used. Moreover, in every grid cell the orientation and inclination of PV panels are assumed to follow Gaussian distributions in which the mean is the optimal value (south orientation and optimal inclination).  Figure \ref{fig_map} depicts the 2015 annually-averaged capacity factors for every country. This approach is conservative, as the time series represent average values per countries but some regions, particularly for north-south oriented countries, could achieve higher capacity factors. \\

The methodology is thoroughly described in \cite{Victoria_2019b} and validated by comparing modelled data with historical values for 15 countries in Europe throughout 2015. The resulting time series are open-licensed and can be retrieved from the repository \href{https://doi.org/10.5281/zenodo.1321809}{10.5281/zenodo.1321809}.\\

Different costs and discount rates are assumed for utility-scale and rooftop installations (Table \ref{tab_costs}). It is assumed that 50\% of the installations would be utility-scale and 50\% on rooftops. As discussed in the section \textit{Limitations of this study}, the impact of this hypothesis is limited. \\

\section{Results} \label{sec_results}

Figure \ref{fig_spatial_plot} depicts the primary energy mix for every country for the optimum power system configuration when the maximum CO$_2$ emissions are limited to 5\%, relative to 1990 level, and the expansion of transmission capacity among neighbouring countries is capped to twice today's volume. Moreover, the average variable renewable energy generation, \textit{i.e.}, solar and wind, is imposed to be proportional to the average electricity demand in every country, that is, countries are to some extent renewable self-sufficient. Under those assumptions, the optimum system comprises large PV capacities in southern countries while, in northern countries, large wind capacities are deployed to exploit the local resource together with reinforced transmission capacities.

\begin{figure*}[!h]
	\centering
	\includegraphics[width=0.9\textwidth]{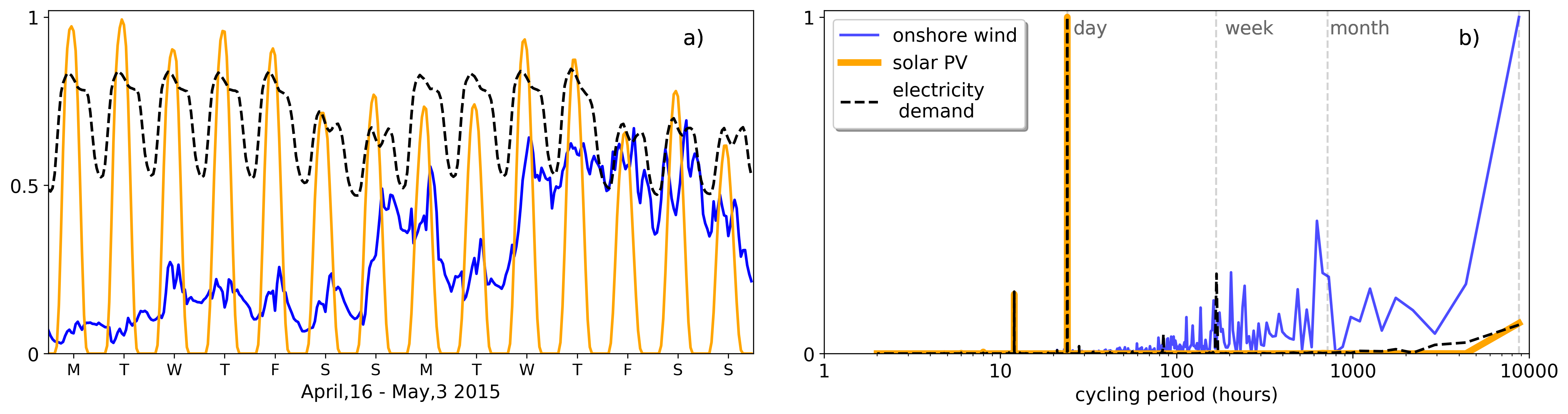}\\
	\caption{(left) Normalised Europe-aggregated electricity demand, solar PV, and onshore wind generation. (right) Fourier power spectra of the time series for the entire year. Vertical grey dashed lines indicate cycling periods corresponding to day, week, month, and year.}
	\label{fig_demand_solar_wind}
\end{figure*}

\begin{figure*} 
	\centering
	\includegraphics[width=0.9\textwidth]{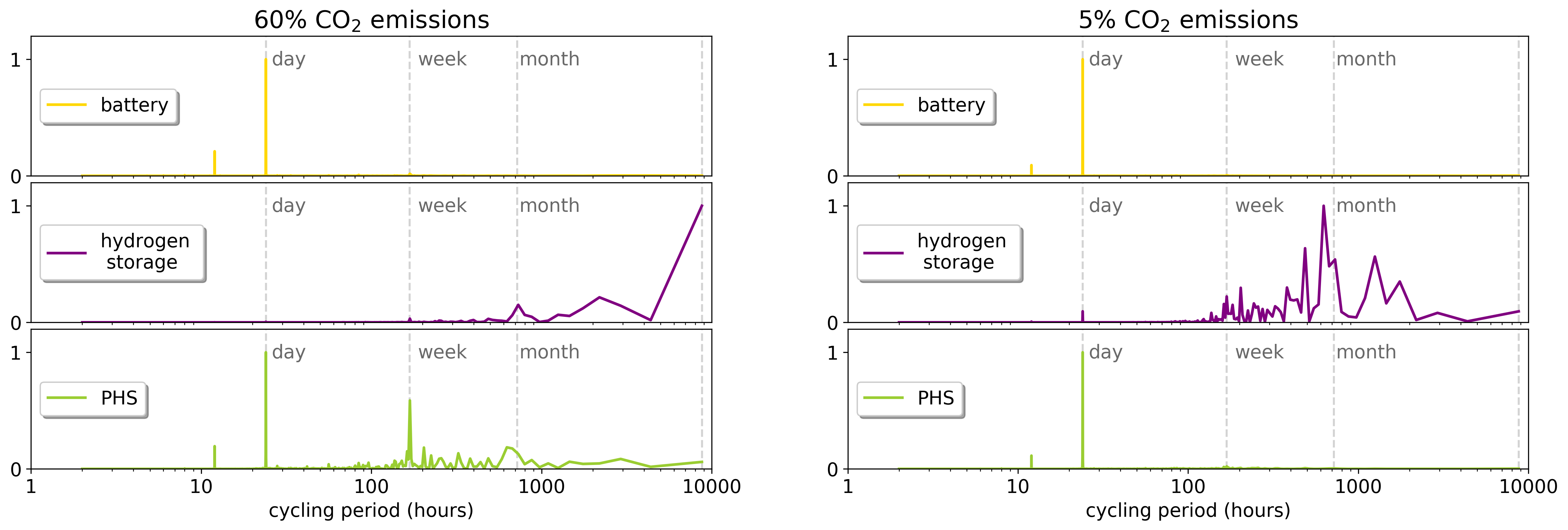}\\
	\caption{Fourier power spectra of the dispatch time series for storage technologies: electric batteries, hydrogen storage, and pumped hydro storage (PHS). 60\% and 5\% CO$_2$ emissions caps are considered. Vertical grey dashed lines indicate cycling periods corresponding to day, week, month, and year.}
	\label{fig_backup_storage}
\end{figure*}

\begin{figure*} 
	\centering
	\includegraphics[width=0.9\textwidth]{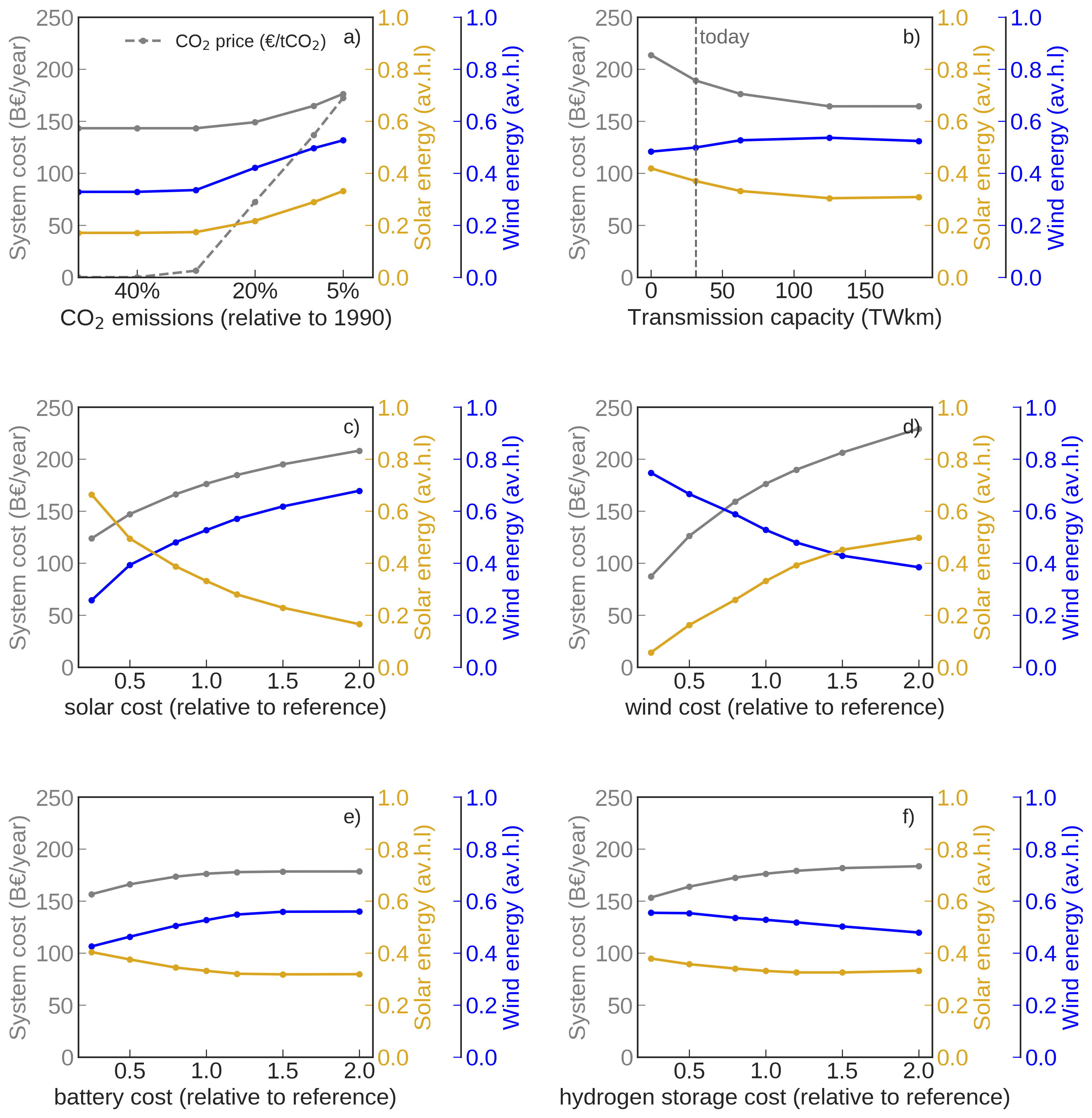}\\
	\caption{Sensitivity of wind and solar electricity generation, and system cost to the main assumptions in the model: (a) CO$_2$ emissions, (b) transmission expansion, (c) solar PV cost, (d) onshore and offshore wind cost, (e) batteries cost, and (f) hydrogen storage cost. }
	\label{fig_sensitivity}
\end{figure*}

\paragraph{Storage} \

Where large solar PV capacities are installed, they are accompanied by significant batteries capacity, see Figure 5 in \cite{Victoria_storage}. Their discharge time at maximum power, which is independently optimised, is in the range of 6 hours to counterbalance the strong daily pattern of PV generation \cite{Victoria_storage}. In countries with high wind penetration, overground hydrogen storage is installed. It has a discharge time of approximately 2 days necessary to balance the synoptic fluctuations of wind generation \cite{Victoria_storage}. The dispatch time series of backup generation and storage technologies are heavily impacted by renewable generation. Figure \ref{fig_demand_solar_wind}(a) shows the Europe-aggregated electricity demand, wind, and solar generation for two weeks in April, 2015. Figure \ref{fig_demand_solar_wind}(b) depicts the Fourier power spectra of the previous time series for the entire year.  Electricity demand shows daily, weakly, and seasonal frequencies; solar PV generation shows daily and seasonal components, while wind generation fluctuates with weakly-monthly and seasonal frequencies. While both wind and solar PV show a seasonal component, their behaviour is opposite. Wind power generation in Europe is much stronger in winter than in summer, and the opposite is true for solar PV \cite{Heide_2010, Heide_2011}. To a certain extent, they can compensate each other, which helps to follow the demand. \\

Figure \ref{fig_backup_storage} depicts the Fourier power spectra for storage dispatch time series when CO$_2$ emissions cap is equal to 60\% and 5\%. As CO$_2$ emissions curb, and consequently renewable penetration increases, the dominant dispatch frequencies of storage technologies increase as shown in \cite{Victoria_storage} and anticipated in \cite{Schlachtberger_2016}. Furthermore, the strong diurnal pattern of solar generation, together with the demand, force the daily frequency on the storage dispatch. Solar generation strongly impacts batteries operation making them charge during the day and discharge throughout the night, see Figure 7 in \cite{Victoria_storage}. Hydrogen storage operation also follows the solar generation, but on the top of that includes several contiguous days of permanent electricity generation through fuel cells or hydrogen production via electrolysis to balance wind fluctuations in the synoptic time scale. For high renewable penetration, the only dominant frequency for PHS is daily, again imposed by PV generation profile.

\paragraph{CO$_2$ emissions limit} \

Figure \ref{fig_sensitivity}(a) depicts the wind and solar PV generation as a function of the maximum CO$_2$ emissions allowed, relative to 1990 level. For  5\% CO$_2$ emissions, the optimum configuration includes an average wind generation equal to 53 \% of the average hourly electricity demand (0.53 av.h.l.), and solar generation equal to 0.33 av.h.l. Optimal installed capacities account for 825 GW of PV, 570 GW of onshore wind, and 59 GW of offshore wind. The remaining energy is provided by hydropower plants, whose capacities are exogenously fixed, and Open-Cycle Gas Turbine (OCGT) power plants. Relaxing the CO$_2$ emissions constraint reduces the installed capacities of both renewable energy sources since more gas can be used to produce electricity.\\

As the PV installed capacity raises, due to its synchronized generation, PV curtailment increases and PV market value gets reduced \cite{Brown_2019}. Market value is defined as the average price of electricity produced by PV relative to the average load-weighted price, and it gives an indication of the value of PV generation in the system. For CO$_2$ emissions below 20\%, the system becomes significantly more expensive \cite{Brown_2019} and large energy capacities for electric batteries and hydrogen storage become cost effective \cite{Victoria_storage}. Restricting even more CO$_2$ emissions cap in the model make technologies such as methanation, \textit{e.g.}, the conversion of direct-air-captured CO$_2$ and hydrogen into methane, competitive. This alternative storage strategy makes use of excess renewable generation reducing PV curtailment and recovering the market value for PV generation \cite{Brown_2019}.\\

Figure \ref{fig_sensitivity}(a) also depicts the shadow price for CO$_2$. This is obtained through the Lagrange/Karush-Kuhn-Tucker multiplier $\mu_{CO2}$ corresponding to the CO$_2$ global cap constraint in the optimisation problem (Eq. \ref{eq_co2cap}). Under our cost assumptions, CO$_2$ constraint is only binding for CO$_2$ emissions below 40\%, in other words, the highly-renewable system is also cheaper without the emissions constraint. This is further discussed later. As CO$_2$ emissions curb, higher CO$_2$ prices are needed to force the emitting technologies out of the system. More detailed analyses on the required CO$_2$ prices for different decarbonisation levels and sectors can be found in \cite{Brown_2018, Zhu_2019, Brown_2019, Victoria_storage}.

\paragraph{Interconnection capacity} \

In Figure \ref{fig_spatial_plot} the transmission expansion is capped to twice today's volume. As this cap is released, Figure \ref{fig_sensitivity}(b) shows how the system evolves towards a more wind-dominated configuration. The reason behind is that increasing transmission allows the spatial smoothing of wind fluctuations in the synoptic scale, reduces the needs for long-term hydrogen storage, and incentivises the installation of large capacities of onshore and offshore wind in countries with better resources. Since PV generation is highly correlated among European countries, increasing the interconnection capacities does not benefit its expansion.\\

The cost decrease attained by expanding interconnections is not linear. For the 30 nodes network, most of the benefits are captured when the transmission capacity is expanded to approximately three times its current volume. Several remarks about this result are relevant. First, as shown in \cite{Brown_2018}, coupling the power system to heating and transport sectors reduces the cost gain provided by transmission expansion. Second, better spatial resolution might be needed to model this effect properly. For the 256 nodes network representing the European power system, capacity expansion equivalent to 1.25 today's value has been shown to be enough to lock in most of the cost benefits of grid expansion \cite{Horsch_2017}. Third, the coarse spatial resolution of the model ignores possible bottleneck in the distribution network neglecting the potential benefits provided by a more distributed PV generation compared to wind. Fourth, interconnection outside of Europe is not included in the model \cite{Dahl_2016}. Expanding transmission capacity to north African countries with high solar resource would probably increase the optimal penetration of solar PV in the network.   

\paragraph{Homogeneous vs. heterogeneous} \

So far, a constraint is included to ensure that the average variable renewable generation in every country is proportional to the electricity demand. When this constraint is released, that is, countries can become net electricity exporter or importer, as a general trend, wind generation increases since higher capacities can be installed  in countries with extremely good resource and exported to neighbours. This is partially because wind resource shows larger differences among neighbouring countries than solar PV.

\paragraph{Costs assumptions} \

Figures \ref{fig_sensitivity}(c)-(f) show the impact of costs assumptions for PV, wind, batteries, and hydrogen storage. Reducing the cost assumed for PV, increases the energy generated by this technology at the expenses of wind and vice versa, but the influence of cost reductions in both technologies are not symmetrical. While very cheap wind pushes PV out of the system almost completely, even under the assumption that PV costs 25\% of the reference values (see Table \ref{tab_costs}), still around 25\% of the demand is covered by wind because higher PV penetrations would require vast amounts of storage that would soar the system cost. Cost reductions for batteries facilitate the integration of PV, increasing the optimal installed capacity of this technology. Cost variations in hydrogen storage have a less noticeable impact on the system configuration. \\

\paragraph{System cost} \

\begin{figure} 
	\centering
	\includegraphics[width=0.8\columnwidth]{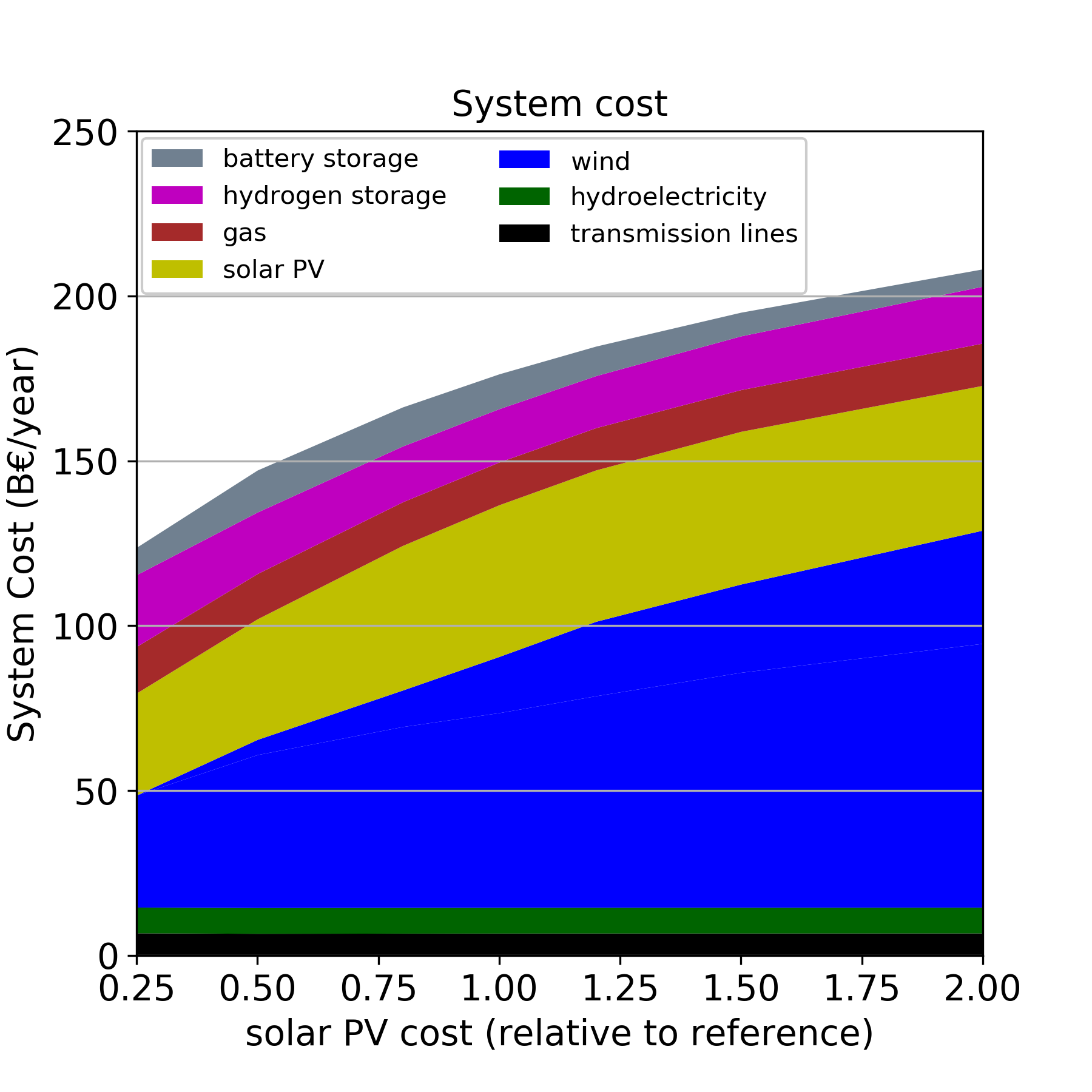}\\
	\caption{Annualised system cost vs. PV cost assumption. Reference values in Table \ref{tab_costs}.}
	\label{fig_system_cost}
\end{figure}

Figure \ref{fig_system_cost} depicts the annualised system cost and its components as a function of the solar PV cost. The main contributions are coming from renewable capacities. As PV cost decreases, lower wind capacity is installed, reducing its contribution to system cost. PV cost contribution is roughly constant because, although higher PV capacity is installed, it has a lower cost. For the reference configuration, the annualised system cost is 176 billion \EUR/year. For today's system, assuming an average cost of electricity of 50 \EUR/MWh, the current power system cost can be estimated in 158 billion \EUR/year \cite{Brown_2018}. Hence, the optimal system configuration fulfilling the 5\% CO$_2$ emissions constraint is only 11 \% more expensive than today.

\paragraph{Sector coupling} \

Coupling the power system with other sectors such as heating, transport, or industry could bring additional flexibility to the system. More detailed results on coupling electricity, heating and transport \cite{Brown_2018, Brown_2019}, the impact of CO$_2$ prices on the electricity-heating system \cite{Zhu_2019, Zhu_sensitivity}, and the role of storage technologies in a sector-coupled European energy system \cite{Victoria_storage} can be found in the provided references. One direct impact of increasing the electrification of transport and heating sectors is that, as electricity demand raises, more solar and wind capacities can be installed before significant storage capacities are needed \cite{Victoria_storage}. For PV, the major impact is observed when the power and transport sectors are coupled. In particular, the scenarios in which Electric Vehicles (EVs) are included, and part of the EV batteries capacity can be used to facilitate the system operation, result in significantly larger PV penetrations in the cost-optimal configurations, at the expenses of the wind generation \cite{Brown_2018, Victoria_storage}. The reason behind is the fact that short-term storage provided by EVs batteries is ideal to counterbalance the strong PV daily generation pattern.\\

Moreover, it is also worth noticing that cooling demand, which is expected to increase due to climate change, is strongly correlated with solar PV generation and makes this technology particularly suitable to supply this demand via heat pumps \cite{Laine_2019, Zhu_sensitivity}.

\paragraph{Limitations of the analysis} \

The PyPSA-Eur-Sec-30 model has some limitations. First, some sectors, such as industry, aviation or shipping are not included. This neglects some possible synergies among them but also the challenges associated with decarbonising those sectors. Some technologies are also lacking, \textit{e.g.} nuclear, biomass, and coal power plants, while storage technologies are limited to electric batteries, hydrogen storage, individual and centralised thermal energy storage. In particular, not including biomass might have a large influence on the CO$_2$ price required to achieve ambitious CO$_2$ reductions, see Figure \ref{fig_sensitivity}(a). The reason is that, in the absence of this dispatchable renewable technology, high CO$_2$ prices are needed to bring into the system large storage and renewable generation capacities, instead of OCGT power plants, to secure the supply of hourly demands. This issue can be especially critical when the heating sector is included, as discussed in \cite{Brown_2018, Victoria_storage}. Second, for the main parameters, sensitivity to cost assumptions is investigated in Figure \ref{fig_sensitivity}, but uncertainties in other assumptions could also affect the results. Third, historical and modelled hourly time series for 2015 are used. It would be desirable to investigate the impact of inter-annual weather variations, as proposed in \cite{Schlachtberger_2017, Collins_2018}. Fourth, while the system configuration is investigated here when CO$_2$ emissions are restricted down to 5\%, it is also interesting to analyze the system under net zero emissions \cite{Brown_2019}. Fifth, PyPSA-Eur-Sec-30 model assumes greenfield optimisation. This is a good approach for the main objectives of this work, \textit{i.e.}, to understand the general system dynamics and obtain inspiring results that can motivate additional and more detailed investigations. However, the power plants currently installed in different European countries and their expected lifetimes will affect the optimal system configuration. For instance, to evaluate the adequate transition path, it is necessary to perform a brownfield optimisation, as in \cite{Breyer_2017}, as well as to consider the implications of using myopic optimisation, without foresight over the investment horizon \cite{Heuberger_2018}, versus optimising the entire transition with perfect foresight.\\

The model limitations that directly impact the optimal PV penetration deserve special mention. First, the coarse-grained network, one node per country, might have important consequences. As the spatial resolution increases, regions with better wind or solar resource can be exploited but also bottlenecks in the transmission and distribution networks become apparent. From the point of view of system cost, in \cite{Horsch_2017}, both effects were found to compensate and roughly constant system cost was obtained when the number of nodes is increased from 37 to 362. Nevertheless, the benefits brought by more distributed rooftop solar PV generation are not adequately captured by this model, since consumers with high electricity tariffs may profit from self-consumption, even though this is not system-optimal. Second, by representing the country-aggregate PV and wind generation through a single node\footnote{For large-area countries, up to 4 different regions are considered to calculate the wind generation profiles. The calculated wind capacities are then joined in a single node, see Section 3.3 in \cite{Schlachtberger_2017}.}, the model favours binary results in some countries, \textit{e.g.} fully wind configuration, when a technology is clearly more competitive. In Figure \ref{fig_spatial_plot}, the optimal system configuration does not include solar PV in Finland, Sweden, Denmark, and Poland although these countries already have a certain PV capacity installed today \cite{EurObservER_2019}. Third, we have assumed that 50\% of the installed PV capacity belongs to utility-scale plants and 50\% to rooftop systems. The costs are different, see Table \ref{tab_costs}, but the discount rates assumed, 7\% for the former and 4\% for the latter, make the annualised costs more similar, 38 and 47 \EUR/kW$\cdot$year for utility-scale and rooftop installations, respectively. If all the PV capacity is assumed to be in the form of utility-scale power plants, annualised PV cost are 11\% lower. Looking at Figure \ref{fig_sensitivity}(c), it can be observed that this would increase the optimal solar PV penetration from 0.33 to 0.36 av.h.l. \\

An alternative approach proposed by Breyer and co-authors consists in independently modelling rooftop PV penetration, based on prosumers behaviour. In \cite{Breyer_2017, Breyer_2017EUPVSEC, Child_2019}, a two steps model is used. First, the prosumers are modelled assuming that they can install their own rooftop PV systems together with Lithium-ion batteries. The prosumers systems are operated to minimise their own electricity cost, calculated as the sum of self-generation and electricity consumed from the grid. The share of prosumers is exogenously fixed in \cite{Breyer_2017EUPVSEC} or limited by a maximum growth in every 5 years step \cite{Child_2019}. In the second step, the capacity and dispatch of prosumers are some of the inputs used to optimise the rest of elements in the energy system.  This two-step method could capture better the deployment of rooftop PV installations, in particular in countries with low solar radiation.  Conversely, in our results, the installed PV capacity, as well as the energy and power capacity of electric batteries in every country, have been optimised simultaneously to the other generation and storage technologies in the system. This global optimisation allows investigating how the optimal capacities and dispatch patterns of the different generation and storage technologies impact each other. \textit{E.g.}, the results indicating that electric batteries capacity and dispatch are heavily influenced by PV generation while hydrogen storage is dimensioned to counterbalance wind fluctuations \cite{Victoria_storage}. Finally, we have assumed that all the PV capacity is static while other configurations that might result cost competitive, such as horizontal 1-axis tracking \cite{Afanasyeva_2018} or delta configuration \cite{Victoria_2019b} are not modelled.

\paragraph{Comparison to similar studies} \

Breyer and co-authors have focused on analysing the role of solar PV in the energy transition \cite{Breyer_2017, Breyer_2017EUPVSEC, Child_2019, Keinera_2019}. For instance, in \cite{Child_2019} two transition paths for Europe are modelled in which generation and storage capacities are optimised in 5-years steps for the period 2015-2050. For the transition path including interconnections among countries, the global average solar PV electricity generation contribution is found to be about 30\% in 2035 when the system achieves zero CO$_2$ emissions and the cost assumptions are similar to those in this paper. This is in agreement with our results in which solar PV generation represents in average 33\% of the electricity demand. Moreover, the higher PV penetration obtained for later years in \cite{Child_2019} can be explained by the lower costs assumed for PV but, in particular, for batteries which incentives the installation of solar capacity at the expenses of wind. There is also a good agreement with the results obtained in the sensitivity analyses for lower PV and battery costs, Figures \ref{fig_sensitivity} (c) and (e).

\section{Conclusions} \label{sec_conclusions}

The role of solar PV in a highly renewable Europe has been investigated by means of an hourly-resolved, one-node-per-country network of the European energy system. For a CO$_2$  emissions constraint equivalent to 5\% of 1990 level, the optimal European power system configuration includes solar PV capacity that generates in average 33\% of the electricity demand. The strong daily pattern of solar PV generation heavily impacts on the dispatch time series of backup generation and storage, in particular, for static electric batteries and pumped hydro storage which show a clearly dominant daily charging and discharging frequency.\\

The optimal penetration of solar PV is influenced by the assumptions within the model, so a sensitivity analysis is carried out to investigate these dependencies. First, as expected, releasing the CO$_2$ emissions constraint reduces the optimal PV capacity since more CO$_2$-emitting electricity can be produced. As CO$_2$ emissions curb and more renewable capacity is installed, curtailment increases and solar PV market value gets reduced, but  the emergence of alternative technologies such as methanation recovers PV market value and reduces curtailment. 
Second, increasing the interconnection capacity among neighbouring countries benefits the installation of wind capacity at the expenses of solar PV. The large correlation among solar generation in different countries, compared to that of onshore and offshore wind, decreases the efficacy of this smoothing strategy for PV. Third, not surprisingly, cost reduction in solar PV or wind increases the optimal capacity of the corresponding technology. However, the effect is not symmetric, and even with very cheap PV, a significant capacity of wind is included in the model. This prevents the installation of huge batteries capacity to counterbalance an only-solar configuration that would soar the system cost. The impact of cost variation in hydrogen storage is small. Conversely, cost decrease in batteries significantly impacts the optimum solar PV capacity on the system. \\

Finally, two effects are particularly relevant for PV when the power system is coupled to other sectors. The first one is related to the possible deployment of electric vehicles whose batteries can charge and discharge into the grid. The additional short-term storage capacity provided by them shifts the optimal configuration towards more solar-dominant. The second one is the fact that cooling demand, expected to increase as a consequence of climate change, correlates extremely well with solar generation providing an opportunity of supplying this demand by solar PV together with heat pumps.

\section{Acknowledgements}

M. Victoria, K. Zhu, G. B. Andresen and M. Greiner are fully or partially funded by the RE-INVEST project, which is supported by  the  Innovation  Fund  Denmark  under  grant  number  6154-00022B. T.B. gratefully acknowledges funding from the Helmholtz Association under grant no. VH-NG-1352. The responsibility for the contents lies solely with the authors.


\end{document}